\documentclass[12pt,a4paper]{article}
\pdfoutput=1
\usepackage[utf8]{inputenc}
\usepackage{amsmath}
\usepackage{amsfonts} 
\usepackage{mathtools} 
\usepackage{amssymb} 
\usepackage{graphicx}
\usepackage{xcolor}
\usepackage{mathrsfs}
\usepackage{hyperref}

\newcommand{\abs}[1]{|#1|} 

\newcommand{\im}[1]{\text{Im}\left(#1\right)}
\newcommand{\refEQ}[1]{eq.\,\eqref{#1}} 
\newcommand{\refEQS}[1]{eqs.\,\eqref{#1}} 

\newcommand{\Yuk}[2]{\Gamma_{#1}^{\rm (#2)}}
\newcommand{\Yukc}[2]{\Gamma_{#1}^{\rm (#2)\ast}}

\newcommand{\Yd}[1]{\Gamma_{#1}^{\rm (d)}}

\newcommand{\Yu}[1]{\Gamma_{#1}^{\rm (u)}}

\newcommand{\Yl}[1]{\Gamma_{#1}^{\rm (e)}}

\newcommand{\Yn}[1]{\Gamma_{#1}^{\rm (\nu)}}

\newcommand{\PR}[1]{{\rm P}_{\!#1}}
\newcommand{\PRX}[2]{P_{#1}^{#2}}

\newcommand{\id}{\mathbf{1}}

\newcommand{\HD}[1]{\Phi_{#1}^{\phantom{\dagger}}}
\newcommand{\HDd}[1]{\Phi_{#1}^\dagger}
\newcommand{\HDC}[1]{\tilde\Phi_{#1}^{\phantom{\dagger}}}

\newcommand{\HHD}[1]{H_{#1}^{\phantom{\dagger}}}

\newcommand{\nHH}{{H}^0}
\newcommand{\nHR}{{R}^0}
\newcommand{\nHI}{{I}^0}
\newcommand{\nh}{{\rm h}}
\newcommand{\nhSM}{{\rm h_{SM}}}
\newcommand{\nH}{{\rm H}}
\newcommand{\nA}{{\rm A}}

\newcommand{\cHp}{{\rm H}^+}
\newcommand{\mnh}{m_{\nh}}

\newcommand{\cb}{c_\beta}
\renewcommand{\sb}{s_\beta}
\newcommand{\tb}{t_\beta}
\newcommand{\tbinv}{\tb^{-1}}

\newcommand{\VEV}[1]{\langle #1 \rangle}
\newcommand{\vev}[1]{v_{#1}}
\newcommand{\ROTmat}{\mathcal R}\newcommand{\ROTmatT}{\ROTmat^T}
\newcommand{\ROT}[1]{\ROTmat_{#1}}
\newcommand{\dCKM}{\delta_{\rm CKM}}
\newcommand{\CKM}{V}\newcommand{\CKMd}{\CKM^\dagger}
\newcommand{\V}[1]{{\CKM_{#1}^{\phantom{\ast}}}}
\newcommand{\Vc}[1]{{\CKM_{#1}^\ast}}
\newcommand{\dPMNS}{\delta_{\rm PMNS}}
\newcommand{\PMNS}{U}\newcommand{\PMNSd}{\PMNS^\dagger}
\newcommand{\U}[1]{{\PMNS_{#1}^{\phantom{\ast}}}}
\newcommand{\Uc}[1]{{\PMNS_{#1}^\ast}}
\newcommand{\OXf}[2]{O_{#1_{#2}}}\newcommand{\OXft}[2]{O_{#1_{#2}}^T}
\newcommand{\OLf}[1]{\OXf{#1}{L}}\newcommand{\OLft}[1]{\OXft{#1}{L}}
\newcommand{\ORf}[1]{\OXf{#1}{R}}\newcommand{\ORft}[1]{\OXft{#1}{R}}
\newcommand{\OLu}{\OLf{u}}\newcommand{\OLut}{\OLft{u}}
\newcommand{\OLd}{\OLf{d}}
\newcommand{\OLn}{\OLf{\nu}}\newcommand{\OLnt}{\OLft{\nu}}
\newcommand{\OLe}{\OLf{e}}\newcommand{\OLet}{\OLft{e}}
\newcommand{\UXf}[2]{U_{#1_{#2}}}\newcommand{\UXfd}[2]{U_{#1_{#2}}^\dagger}
\newcommand{\ULf}[1]{\UXf{#1}{L}}\newcommand{\ULfd}[1]{\UXfd{#1}{L}}
\newcommand{\ULud}{\ULfd{u}}
\newcommand{\ULd}{\ULf{d}}
\newcommand{\ULn}{\ULf{\nu}}
\newcommand{\ULed}{\ULfd{e}}
\newcommand{\roto}[2]{{\rm R}_{#1}(#2)}
\newcommand{\rot}[2]{R_{#1}(#2)}\newcommand{\rott}[2]{R_{#1}(#2)^T}
\newcommand{\urvec}[2]{\hat r_{[#1]#2}}

\newcommand{\phase}[2]{\varphi_{#1}(#2)}
\newcommand{\mMQ}[1]{M_{#1}^{\phantom{\dagger}}}

\newcommand{\wMQ}[1]{M_{#1}^{0}}\newcommand{\wMQd}[1]{M_{#1}^{0\dagger}}
\newcommand{\wMU}{\wMQ{u}}
\newcommand{\wMD}{\wMQ{d}}
\newcommand{\wMN}{\wMQ{\nu}}
\newcommand{\wML}{\wMQ{\ell}}
\newcommand{\hwMQ}[1]{\widehat M_{#1}^{0}}\newcommand{\hwMQt}[1]{\widehat M_{#1}^{0\,T}}
\newcommand{\mNQ}[1]{N_{#1}^{\phantom{\dagger}}}

\newcommand{\wNQ}[1]{N_{#1}^{0}}
\newcommand{\wNU}{\wNQ{u}}
\newcommand{\wND}{\wNQ{d}}
\newcommand{\wNN}{\wNQ{\nu}}
\newcommand{\wNL}{\wNQ{\ell}}
\newcommand{\Zn}[1]{\mathbb{Z}_{#1}}
\newcommand{\ZZ}{\Zn{2}}

\graphicspath{{Figs/}}

\begin{document}

\hfill\begin{minipage}[r]{0.3\textwidth}\begin{flushright}  CFTP/21-08\\ IFIC/21-17 \end{flushright} \end{minipage}

\begin{center}

\vspace{0.50cm}

{\large \bf {The framework for a common origin of $\dCKM$ and $\dPMNS$}}\\
\vspace{0.50cm}

Joao M. Alves$^{a}$\footnote{\texttt{j.magalhaes.alves@tecnico.ulisboa.pt}},
Francisco J. Botella$^{b}$\footnote{\texttt{Francisco.J.Botella@uv.es}},
Gustavo C. Branco$^{a}$\footnote{\texttt{gbranco@tecnico.ulisboa.pt}},\\ 
Fernando Cornet-Gomez$^{b}$\footnote{\texttt{Fernando.Cornet@ific.uv.es}},
Miguel Nebot$^{b}$\footnote{\texttt{Miguel.Nebot@uv.es}}
\end{center}
 
\vspace{0.50cm}
\begin{flushleft}
\emph{$^a$ CFTP, Instituto Superior T\'ecnico, U. de Lisboa,\\ 
\quad Av. Rovisco Pais 1, P-1049-001 Lisboa, Portugal.}\\
\emph{$^b$ Departament de F\`\i sica Te\`orica and Instituto de F\' \i sica Corpuscular (IFIC),\\
\quad Universitat de Val\`encia -- CSIC, E-46100 Valencia, Spain.}
\end{flushleft}

\begin{abstract}
We analyse a possible connection between CP violations in the quark and lepton sectors, parametrised by the CKM and PMNS phases. If one assumes that CP breaking arises from complex Yukawa couplings, both in the quark and lepton sectors, the above connection is not possible in general, since Yukawa couplings in the two sectors have independent flavour structures. We show that both the CKM and PMNS phases can instead be generated by a vacuum phase in a class of two Higgs doublet models, and in this case a connection may be established. This scenario requires the presence of scalar FCNC at tree level, both in the quark and lepton sectors. The appearance of these FCNC is an obstacle and a blessing. An obstacle since one has to analyse which models are able to conform to the strict experimental limits on FCNC, both in the quark and lepton sectors. A blessing, because this class of models is falsifiable since FCNC arise at a level which can be probed experimentally in the near future, specially in the processes $\nh\to e^\pm\tau^\mp$ and $t\to\nh c$. The connection between CP violations in CKM and PMNS is explicitely illustrated in models with Minimal Flavour Violation.

\end{abstract}

 \clearpage
\section{Introduction\label{SEC:Introduction}}
In this paper we study the possibility of having a relation between CP violation in the quark and lepton sectors, parametrized by the phases $\dCKM$ and $\dPMNS$,  respectively, of the CKM \cite{Cabibbo:1963yz,Kobayashi:1973fv} and PMNS \cite{Pontecorvo:1957qd,Maki:1962mu} mixing matrices. This question is specially important in view of the efforts to detect leptonic CP violation in neutrino experiments such as Dune, T2K and Nova \cite{Acciarri:2016crz,Abe:2019vii,Acero:2019ksn}. At present, there is solid experimental evidence that CKM is complex, even if one allows for the presence of New Physics contributing to CP violation \cite{Botella:2005fc,Bona:2005eu}.
The fact that CKM is complex does not imply that CP violation is violated at the Lagrangian level through complex Yukawa couplings. Indeed it has been pointed out \cite{Nebot:2018nqn} that one may have a vacuum induced CP violation generating a complex CKM matrix in agreement with experiment. If one considers an extension of the Standard Model (SM) with non-vanishing neutrino masses and assumes that the origin of CP violation is the presence of complex Yukawa couplings, then there is no relation between $\dCKM$ and $\dPMNS$. This just reflects the fact that in the SM the lepton and the quark Yukawa couplings are completely independent quantities. If one wants to obtain a relation between $\dCKM$ and $\dPMNS$, an interesting possibility is to assume that CP is spontaneously broken, with the vacuum phase generating both the phase in the quark and in the lepton sectors. In this paper, we consider the extension to the leptonic sector of a previously proposed viable minimal model where the Lagrangian respects CP invariance but the vacuum is CP violating, with a complex phase which generates a complex CKM matrix. The model consists of a generalised Branco-Grimus-Lavoura (BGL) model \cite{Branco:1996bq,Botella:2009pq,Botella:2011ne,Alves:2017xmk} with a flavoured $\ZZ$ symmetry in the context of two Higgs doublet models (2HDMs) \cite{Branco:2011iw,Ivanov:2017dad}. Two of the quark families are even under the $\ZZ$ symmetry while the remaining one is odd.
It was shown \cite{Nebot:2018nqn} that there is a profound connection between the possibility of generating a complex CKM matrix and the existence of tree level Scalar Flavour Changing Neutral Couplings (SFCNC) both in the up and down quark sectors. The same correlation also occurs when one extends the model to the leptonic sector. In order for $\dPMNS\neq 0$ to be generated one has to have SFCNC both in the charged lepton and neutrino sectors. The fact that SFCNC necessarily arise in the class of models which we are considering is both an obstacle and a blessing. It is an obstacle because one has to analyse which models are able to conform to the strict experimental limits on SFCNC in the quark and lepton sectors. It is a blessing because the class of models we are considering are falsifiable in the sense that they imply SFCNC at a level which can be probed at different experiments. 
The paper is organised as follows. In the next section we present the model, including the quark and lepton sectors. In section \ref{SEC:CPgeneration} we analyse how $\dCKM$ and $\dPMNS$ are generated. In section \ref{SEC:CPSFCNC} we analyse how CP violation requires the presence of SFCNC. Section \ref{SEC:genrel} is dedicated to a study of a general relation between $\dCKM$ and $\dPMNS$. In section \ref{SEC:simpMFV} we present simplified models incorporating Minimal Flavour Violation (MFV) \cite{DAmbrosio:2002vsn,Cirigliano:2005ck} and the connection to SFCNC; a specific model is analysed in detail, including the prediction of $\dPMNS$ arising in this subclass of models. Finally in the last section, we present our conclusions. Additional details on different aspects are provided in the appendices. 

\section{The model\label{SEC:Model}}
The Yukawa sector of the most general 2HDM, in the case of Dirac neutrinos, can be written as
\begin{equation}\label{eq:YukLag:00}
\mathscr L_{Y}=-\sum_{i=1}^{2}\Big[\,
\overline{Q_{L}^{0}} \,\Yd{i}\HD{i}\,d_{R}^{0}+\overline{Q_{L}^{0}}\,\Yu{i}\HDC{i}\, u_{R}^{0}+
\overline{L_{L}^{0}} \,\Yl{i}\HD{i}\,e_{R}^{0}+\overline{L_{L}^{0}}\,\Yn{i}\HDC{i}\,\nu_{R}^{0}\,
\Big]+\text{h.c.}\,.
\end{equation}
We extend the viable model presented in reference \cite{Nebot:2018nqn} to the leptonic sector. The model is enforced by a $\ZZ$ symmetry, extended now to the leptonic sector. The fields $\HD{2}$, $Q_{L_{3}}^0$ and $L_{L_{3}}^0$ are odd under $\ZZ$, the rest of the fields are even under $\ZZ$. This symmetry gives rise to the so-called generalized BGL (gBGL) textures \cite{Alves:2017xmk} for the original Yukawa coupling matrices $\Yuk{i}{f}$:
\begin{eqnarray}
\label{eq:textures:1}
&\Yd{1} \sim \Yu{1} \sim \Yl{1} \sim \Yn{1} \sim
\begin{pmatrix}\times & \times & \times \\ \times & \times & \times \\ 0 & 0 & 0\end{pmatrix}, \\
\label{eq:textures:2}
&\Yd{2} \sim \Yu{2} \sim \Yl{2} \sim \Yn{2} \sim
\begin{pmatrix}0 & 0 & 0 \\ 0 & 0 & 0 \\ \times & \times & \times\end{pmatrix},%
\end{eqnarray}
with $\times$ denoting generic entries. \\
Note that $\Yuk{2}{f}=\PR{3}\Yuk{2}{f}$ and $\Yuk{1}{f}=\left(\id-\PR{3}\right) \Yuk{1}{f}$ where $\PR{3}$ is the projector 
\begin{equation}
\PR{3}=
\begin{pmatrix}
0 & 0 & 0 \\ 0 & 0 & 0 \\ 0 & 0 & 1%
\end{pmatrix}.%
\label{projector}
\end{equation}%
Spontaneous electroweak symmetry breaking is achieved with the vacuum
\begin{equation}
\VEV{\HD{j}}=\frac{\vev{j}e^{i\theta_j}}{\sqrt{2}}\begin{pmatrix}0\\ 1\end{pmatrix},\quad
\vev{j},\theta_j\in\mathbb{R},\ \vev{j}>0.
\end{equation}
As usual, we introduce $\vev{}^2\equiv\vev{1}^2+\vev{2}^2$ and $\beta$ such that $\vev{1}=\vev{}\cb$, $\vev{2}=\vev{}\sb$ (here and in the following, $\cb=\cos\beta$, $\sb=\sin\beta$, $\tb=\tan\beta$, $\tbinv=(\tb)^{-1}$) and impose $\vev{}^2=\frac{1}{\sqrt{2}G_F}\simeq(246\text{ GeV})^2$. 
Instead of $\HD{j}$, $\HHD{1}$ and $\HHD{2}$ such that
\begin{equation}
\VEV{\HHD{1}}=\frac{\vev{}}{\sqrt{2}}\begin{pmatrix}0\\ 1\end{pmatrix},\qquad
\VEV{\HHD{2}}=\begin{pmatrix}0\\ 0\end{pmatrix},
\end{equation}
are introduced; $\{\HHD{1},\HHD{2}\}$ is the ''Higgs basis'' \cite{Georgi:1978ri,Botella:1994cs}, that is
\begin{equation}
\begin{pmatrix}\HHD{1}\\ \HHD{2}\end{pmatrix}=
\begin{pmatrix}\cb& \sb\\ \sb& -\cb\end{pmatrix}
\begin{pmatrix}e^{-i\theta_1}\HD{1}\\ e^{-i\theta_2}\HD{2}\end{pmatrix}.
\end{equation}
One can rewrite the Yukawa couplings in \refEQ{eq:YukLag:00} as
\begin{eqnarray}
-\mathscr L_{Y} &=&
\overline{Q^{0}_{L}}\frac{\sqrt{2}}{\vev{}}\left[\wMD H_{1}+\wND H_{2}\right] d_{R}^{0}+\overline{Q^{0}_{L}}\frac{\sqrt{2}}{\vev{}}\left[\wMU \widetilde{H}_{1}+\wNU\widetilde{H}_{2}\right] u_{R}^{0}+  \label{Quark Yukawa} \\
&&\overline{L^{0}_{L}}\frac{\sqrt{2}}{\vev{}}\left[\wML H_{1}+\wNL H_{2}\right] e_{R}^{0}+\overline{L^{0}_{L}}\frac{\sqrt{2}}{\vev{}}\left[\wMN\widetilde{H}_{1}+\wNN\widetilde{H}_{2}\right] \nu _{R}^{0}+\text{h.c.}  \label{lepton Yukawa}
\end{eqnarray}%
where
\begin{equation}
\begin{aligned}
\wMD=\frac{e^{i\theta_{1}}\vev{}}{\sqrt{2}}\left(\Yd{1}\cb+\Yd{2}\sb e^{i\theta}\right), &&
\wMU=\frac{e^{-i\theta_{1}}\vev{}}{\sqrt{2}}\left(\Yu{1}\cb+\Yu{2}\sb e^{-i\theta}\right), \\ 
\wML=\frac{e^{i\theta_{1}}\vev{}}{\sqrt{2}}\left(\Yl{1}\cb+\Yl{2}\sb e^{i\theta}\right),&& 
\wMN=\frac{e^{-i\theta_{1}}\vev{}}{\sqrt{2}}\left(\Yn{1}\cb+\Yn{2}\sb e^{-i\theta}\right),
\end{aligned}
\label{Mass matrices}
\end{equation}
with $\theta=\theta_{2}-\theta_{1}$.%
The corresponding expressions for $\wNQ{f}$ in terms of $\Yuk{i}{f}$, $\beta$ and $\theta_{i}$ are
\begin{equation}\label{eq:N:matrices:00}
\begin{aligned}
\wND=\frac{e^{i\theta_{1}}\vev{}}{\sqrt{2}}\left(\Yd{1}\sb-\Yd{2}\cb e^{i\theta}\right), &&
\wNU=\frac{e^{-i\theta_{1}}\vev{}}{\sqrt{2}}\left(\Yu{1}\sb-\Yu{2}\cb e^{-i\theta}\right), \\ 
\wNL=\frac{e^{i\theta_{1}}\vev{}}{\sqrt{2}}\left(\Yl{1}\sb-\Yl{2}\cb e^{i\theta}\right),&& 
\wNN=\frac{e^{-i\theta_{1}}\vev{}}{\sqrt{2}}\left(\Yn{1}\sb-\Yn{2}\cb e^{-i\theta}\right).
\end{aligned}
\end{equation}%
The most relevant property is the following:
\begin{equation}\label{SFCNC}
\wNQ{f}=\left[ \tb\id-\left(\tb+\tbinv\right)\PR{3}\right]\wMQ{f}\,.  
\end{equation}%
Note that even if $\wNQ{f}$ are proportional to $\wMQ{f}$, this proportionality involves a diagonal matrix different from the identity. This means that in general it will not be possible to bi-diagonalize both matrices simultaneously. The matrices $\wNQ{f}$ control the scalar mediated flavour changing neutral couplings.\\ 
Concerning the scalar sector, the scalar potential has the following form
\begin{multline}\label{eq:ScalarPotential}
V(\HD{1},\HD{2})=
\mu_{11}^2\HDd{1}\HD{1}+\mu_{22}^2\HDd{2}\HD{2}+\mu_{12}^2\left(\HDd{1}\HD{2}+\HDd{2}\HD{1}\right)
+\lambda_1(\HDd{1}\HD{1})^2+\lambda_2(\HDd{2}\HD{2})^2\\+2\lambda_3(\HDd{1}\HD{1})(\HDd{2}\HD{2})+2\lambda_4(\HDd{1}\HD{2})(\HDd{2}\HD{1})
+\lambda_5\left((\HDd{1}\HD{2})^2+(\HDd{2}\HD{1})^2\right)\,,
\end{multline}
with $\mu_{jk}^2\in\mathbb{R}$ and $\lambda_k\in\mathbb{R}$. For a scalar potential with an exact $\ZZ$ symmetry, $\mu_{12}^2=0$ and even if $\lambda_5$ could be complex, there is no CP violation. In \refEQ{eq:ScalarPotential}, $\mu_{12}^2\neq 0$ softly breaks the $\ZZ$ symmetry and even though $\mu_{12}^2,\lambda_5\in\mathbb{R}$ give a CP invariant $V(\HD{1},\HD{2})$, the possibility to have CP violation arising from the vacuum is open. The expansion of the fields around the vacuum in the Higgs basis reads
\begin{equation}\label{eq:Fields:HB}
\HHD{1} = \begin{pmatrix} G^+ \\ \frac{v+\nHH+iG^0}{\sqrt{2}} \end{pmatrix},\quad 
\HHD{2} = \begin{pmatrix} \cHp \\ \frac{\nHR+i\nHI}{\sqrt{2}} \end{pmatrix}.
\end{equation}
The neutral mass eigenstates are $\{\nh,\nH,\nA\}$:
\begin{equation}\label{eq:ScalarMass}
\begin{pmatrix} \nh\\ \nH\\ \nA \end{pmatrix}=
\ROTmatT
\begin{pmatrix} \nHH\\ \nHR\\ \nHI \end{pmatrix}\,,
\end{equation}
with $\ROTmat$ a real orthogonal $3\times 3$ matrix. $\nh$ is usually assumed to be the SM-like Higgs with mass $\mnh=125$ GeV; the alignment limit, in which its couplings are SM-like, corresponds to $\ROT{11}\to 1$. For further details concerning the scalar sector we refer to \cite{Nebot:2018nqn}.

\section{Generation of CP violating CKM and PMNS matrices\label{SEC:CPgeneration}}
It is clear that the global phase $\theta_{1}$ in the previous expressions can be rephased away by redefining the phases of the $\HHD{i}$ fields or of the right handed fermion fields, and thus we set $\theta_{1}=0$ from now on without loss of generality. Note also that invariance under CP of the entire Lagrangian implies that 
\begin{equation}\label{eq:realYuk}
\Yuk{i}{f}=\Yukc{i}{f}\,.
\end{equation}%
As shown in reference \cite{Nebot:2018nqn}, taking into account the reality condition in \refEQ{eq:realYuk} and the position of the irremovable phase $\theta$ in the textures in \refEQS{eq:textures:1}-\eqref{eq:textures:2}, the mass matrices can be factorized as
\begin{equation}\label{eq:MassMatrixStruct}
\wMQ{f}=
\begin{pmatrix}1 & 0 & 0 \\ 0 & 1 & 0 \\ 0 & 0 & e^{i\sigma_{f}}\end{pmatrix}%
\hwMQ{f}=\phase{3}{\sigma_f}\,\hwMQ{f}\,,
\end{equation}%
where $\hwMQ{f}$ are arbitrary real mass matrices and $\sigma_d=\sigma_e=\theta$, $\sigma_u=\sigma_\nu=-\theta$. Note that the diagonal matrix of phases $\phase{3}{\sigma_f}$ can be written as
\begin{equation}\label{eq:PhaseProyector}
\phase{3}{\sigma_f}=\id+(e^{i\sigma_{f}}-1)\PR{3}.
\end{equation}
It is clear that $\theta$, the relative phase among the vacuum expectation values of the original scalar fields, is the unique source of irremovable complexity in the mass matrices, and thus it is the candidate to generate the phases in the CKM and PMNS mixing matrices. To analyse the mixing matrices it is sufficient to diagonalize $\wMQ{f}\wMQd{f}$: following \refEQ{eq:MassMatrixStruct},
\begin{equation}
\wMQ{f}\wMQd{f}=\phase{3}{\sigma_{f}}\hwMQ{f}\hwMQt{f}\phase{3}{-\sigma_{f}}.
\label{mass square left}
\end{equation}%
Since $\hwMQ{f}\hwMQt{f}$ is real, symmetric (and positive definite), it can be diagonalized with a real orthogonal (rotation) matrix $\OLf{f}$%
\begin{equation}
\OLft{f}\hwMQ{f}\hwMQt{f}\OLf{f}=\text{diag}(m_{f_{1}}^{2},m_{f_{2}}^{2},m_{f_{3}}^{2}).
\label{mass square left diagonalization}
\end{equation}%
One obtains trivially
\begin{equation}\label{left rotation}
\ULfd{f}\wMQ{f}\wMQd{f}\ULf{f}=\text{diag}(m_{f_{1}}^{2},m_{f_{2}}^{2},m_{f_{3}}^{2}),\quad \ULf{f}=\phase{3}{\sigma_{f}}\OLf{f}.
\end{equation}
Since $\wMQd{f}\wMQ{f}=\hwMQt{f}\hwMQ{f}$ is real and symmetric, we will also have 
\begin{equation}\label{mass square right diagonalization}
\ORft{f}\hwMQt{f}\hwMQ{f}\ORf{f}=\text{diag}(m_{f_{1}}^{2},m_{f_{2}}^{2},m_{f_{3}}^{2}),
\end{equation}%
in such a way that the fermion mass matrix bi-diagonalization reads
\begin{equation}\label{Mass matrices diagonalization}
\mMQ{f}=\ULfd{f}\wMQ{f}\ORf{f}=
\begin{pmatrix}m_{f_{1}} & 0 & 0 \\ 0 & m_{f_{2}} & 0 \\ 0 & 0 & m_{f_{3}}\end{pmatrix}.%
\end{equation}%
Correspondingly, the CKM mixing matrix $\CKM=\ULud\ULd$ and the PMNS mixing matrix $\PMNS=\ULed\ULn$, defined with the usual conventions in the charged currents $W^\pm$ interactions, are:%
\begin{equation}\label{CKM and PMNS}
\CKM=\OLut\,\phase{3}{2\theta}\,\OLd,\quad \PMNS=\OLet\,\phase{3}{-2\theta}\,\OLn\,.
\end{equation}%
Since $\OLf{f}$ are arbitrary real rotations, it is evident that there is enough freedom in \refEQ{CKM and PMNS} to obtain arbitrary $\CKM$ and $\PMNS$, except for the fact that any CP violating observable in the quark sector and any CP violating observable in the lepton sector, must vanish with $\theta\to 0$. It is thus interesting to scrutinize in detail the relation that must exist among the CP violating phases in $\CKM$ and $\PMNS$, $\dCKM$ and $\dPMNS$ respectively. Anticipating the discussion in section \ref{SEC:genrel}, $\dCKM$ and $\dPMNS$ will simply correspond to the CP phases in a standard parametrization; notice that, a priori, the change of sign in $\theta$ entering $\CKM$ and $\PMNS$ in \refEQ{CKM and PMNS}, does not imply in general that $\dCKM=-\dPMNS$.

\section{CP violation and the presence of SFCNC\label{SEC:CPSFCNC}}
It was shown in \cite{Nebot:2018nqn} that in this class of 2HDMs with spontaneous CP violation, there is a deep connection between the complexity of the CKM matrix and the presence of SFCNC. Since there is no evidence yet of SFCNC beyond the SM, the simplest approach in the analysis of these models would be to impose that SFCNC are absent. As discussed in \cite{Nebot:2018nqn}, this leads to a real CKM, contrary to evidence, and thus SFCNC are necessary. We recall here the essence of this connection.\\ 
The appearance of SFCNC is encoded in the $\wNQ{f}$ matrices in \refEQ{eq:N:matrices:00}, which control the Yukawa couplings of $\HHD{2}$; in the fermion mass bases, $\wNQ{f}\to\mNQ{f}$, and \refEQ{SFCNC} gives 
\begin{multline}\label{SFCNC_Mass_Basis}
\mNQ{f}=\ULfd{f}\,\wNQ{f}\,\ORf{f}=\left[\tb\id-\left(\tb+\tbinv\right)\PRX{3}{f}\right]\mMQ{f}\\
=\left[\tb\id-\left(\tb+\tbinv\right)\PRX{3}{f}\right]\text{diag}(m_{f_1},m_{f_2},m_{f_3}),
\end{multline}%
where we have introduced the projection operators 
\begin{equation}\label{fermion projectors}
\PRX{3}{f}\equiv\ULfd{f}\,\PR{3}\,\ULf{f}=\OLft{f}\,\PR{3}\,\OLf{f}.
\end{equation}%
In \refEQS{SFCNC_Mass_Basis}-\eqref{fermion projectors}, SFCNC are controlled by the real projectors $\PRX{3}{f}$, in particular the off-diagonal entries of $\PRX{3}{f}$, which are controlled by the $\OLf{f}$ matrices, which also give the CKM and PMNS mixing matrices.\\
It is important to notice that, by construction, 
\begin{equation}\label{fermion projector relations}
\PRX{3}{u}=\CKM\,\PRX{3}{d}\,\CKMd,\quad \PRX{3}{e}=\PMNS\,\PRX{3}{\nu}\,\PMNSd.
\end{equation}%
Equation \eqref{fermion projector relations} means that SFCNC in the up and down quark sectors are not independent, they are related through the CKM matrix. For example, if one fixes SFCNC in the up quark sector, SFCNC in the down quark sector are completely determined; this fact will be particularly relevant in order to address appropriately the count and the election of the independent parameters in the model.
The situation in the lepton sector is analogous.\\ The elements of the matrices $\PRX{3}{f}$ are
\begin{equation}
\left(\PRX{3}{f}\right)_{ij}=\left(\OLft{f}\,\PR{3}\,\OLf{f}\right)_{ij}=
\left(\OLf{f}\right)_{3i}\left(\OLf{f}\right)_{3j}\equiv 
\urvec{f}{i}\urvec{f}{j}\,,
\label{projectors_from_r's}
\end{equation}%
where $\urvec{f}{i}\equiv\left(\OLf{f}\right)_{3i}$ are the components of real, unit vectors in three dimensions $\urvec{f}{}$, the third rows of the orthogonal matrices $\OLf{f}$. In principle each $\urvec{f}{}$ would require two independent parameters, but it follows from \refEQ{fermion projector relations} that $\urvec{u}{j}\V{jk}=e^{2i\theta }\urvec{d}{k}$ and $\urvec{e}{j}\U{jk}=e^{-2i\theta}\urvec{\nu}{k}$ in such a way that $\urvec{d}{k}$ is fixed once $\urvec{u}{j}$ and $\V{jk}$ are known, and similarly for $\urvec{\nu}{k}$ with respect to $\urvec{e}{j}$ and $\U{jk}$. The only way to avoid SFCNC in $\PRX{3}{f}$ is to set one component $\urvec{f}{k}=1$ and the others $\urvec{f}{j}=0$, $j\neq k$. In that case
\begin{equation}
\left(\PRX{3}{f}\right)_{ij}=\delta_{ik}\delta_{jk}\equiv \left(
\PR{k}\right)_{ij}  \label{diagonal projectors}
\end{equation}
for a fixed $k$, i.e. $\PRX{3}{f}=\PR{k}$ for that given $f$. Consider, for example, the absence of SFCNC in the neutrino sector, that is $\PRX{3}{\nu}=\PR{k}$. Using \refEQ{eq:PhaseProyector} in \refEQ{CKM and PMNS},
\begin{equation}\label{eq:SFCNC:CPV:nu:1}
\begin{aligned}
\PMNS &=\OLet\phase{3}{-2\theta}\OLn=
\OLet\OLn\left[\id+(e^{-2i\theta}-1)\OLnt\PR{3}\OLn\right]\\
&=\OLet\OLn\left[\id+(e^{-2i\theta}-1)\PRX{3}{\nu}\right]=\OLet\OLn\left[\id+(
e^{-2i\theta}-1)\PR{k}\right].
\end{aligned}
\end{equation}
Then the PMNS matrix $\PMNS$ is written as a real rotation times a diagonal matrix of phases (with $e^{-2i\theta}$ in position $k$ and the rest of them $1$), and thus there is no CP violation. Similarly, if one starts with the absence of SFCNC in the charged lepton sector, $\PRX{3}{e}=\PR{k}$ and
\begin{equation}\label{eq:SFCNC:CPV:nu:2}
\begin{aligned}
\PMNS &=\OLet\phase{3}{-2\theta}\OLn=
\left[\id+(e^{-2i\theta}-1)\OLet\PR{3}\OLe\right]\OLet\OLn\\
&=\left[\id+(e^{-2i\theta}-1)\PRX{3}{e}\right]\OLet\OLn=\left[\id+(
e^{-2i\theta}-1)\PR{k}\right]\OLet\OLn,
\end{aligned}
\end{equation}
giving again a CP conserving mixing matrix. Therefore, in this model, in order to have a non-vanishing CP violating phase in the CKM matrix, there must be tree level SFCNC both in the up and in the down quark sectors and, mutatis mutandis, in order to have a non-vanishing CP violating phase in the PMNS matrix, there must be tree level SFCNC both in the neutrino and in the charged lepton sectors.

\section{The general relation between $\dCKM$ and $\dPMNS$\label{SEC:genrel}}
From the discussion in the previous sections, it is important to recall that 
\begin{itemize}
\item[(i)] $\theta\neq 0$ arising from the vacuum is the only possible source of CP violation in the CKM and in the PMNS mixing matrices, 
\item[(ii)] if SFCNC are removed in one fermion sector, CP violation in the corresponding mixing matrix disappears even if $\theta\neq 0$.
\end{itemize}
The CKM and PMNS matrices can be parametrized, up to rephasing of fields, in terms of 4 quantities each; in the usual PDG parametrizations \cite{Zyla:2020zbs}, the parameters are $\{\theta_{12}^q,\theta_{13}^q,\theta_{23}^q,\delta_q\}$ and $\{\theta_{12}^\ell,\theta_{13}^\ell,\theta_{23}^\ell,\delta_\ell\}$ respectively. In the quark sector, experimental information allows the extraction of $\theta_{12}^q$, $\theta_{13}^q$, $\theta_{23}^q$, and of the CP violating phase $\delta_q$, which is neatly different from zero. In the lepton sector, experimental information allows to extract $\theta_{12}^\ell$, $\theta_{13}^\ell$, $\theta_{23}^\ell$, but the phase $\delta_\ell$ remains 'the last frontier', where some sensitivity is emerging in current analyses \cite{deSalas:2020pgw,Esteban:2020cvm}, but still far from a neat determination.\\ 
Although the form of the CKM and PMNS matrices in \refEQ{CKM and PMNS} is different from the PDG parametrization, one can impose in an invariant manner that $\CKM$ and $\PMNS$ agree with the experimental information encoded in $\{\theta_{12}^q,\theta_{13}^q,\theta_{23}^q,\delta_q\}$ and $\{\theta_{12}^\ell,\theta_{13}^\ell,\theta_{23}^\ell,\delta_\ell\}$ (see appendix \ref{SEC:CKMPMNSfits} for a detailed explanation). In particular, $\dCKM$, the CP violating phase in $\CKM$, is the model prediction for $\delta_q$ and similarly $\dPMNS$, the CP violating phase in $\PMNS$, is the model prediction for $\delta_\ell$. We already know that if $\theta=0$, then $\dCKM=\dPMNS=0$, but $\theta\neq 0$ $\nRightarrow$ $\dCKM\neq 0$, $\dPMNS\neq 0$.\\ 
At this point, we need to discuss the independent parameters in the model. We start with the quark sector, keeping in mind that an important goal is to analyse how information on CP violation in the quark sector, i.e. the constraint $\delta_q=\dCKM$, can translate into some prediction on $\dPMNS$.\\ 
The CKM matrix in \refEQ{CKM and PMNS} involves 7 real parameters: 3 in the rotation $\OLu$, 3 in the rotation $\OLd$, and $\theta$. It should match the four independent quantities measured in the CKM matrix, equivalent to $\{\theta_{12}^q,\theta_{13}^q,\theta_{23}^q,\delta_q\}$. Besides the CKM matrix, the parameters in $\OLu$ and $\OLd$, in particular the ones in $(\OLu)_{3j}=\urvec{u}{j}$, $(\OLd)_{3j}=\urvec{d}{j}$ (see \refEQ{projectors_from_r's}) also control the SFCNC which, we recall, must be present: they involve 2 of the parameters in $\OLu$, $\OLd$. This means that measurements of CKM and SFCNC processes (e.g. $\nh\to u\bar c,\bar u c$, $t\to\nh c$, $t\to\nh u$, $\nh\to b\bar s,\bar bs$, etc), that is measurements of $\{\theta_{12}^q,\theta_{13}^q,\theta_{23}^q,\delta_q\}$ and of $\{\urvec{u}{1},\urvec{u}{2}\}$ (or, equivalently, of 2 independent $\urvec{u}{j},\urvec{d}{k}$), can provide sufficient constraints to fix the parameters of the model. Although there are in principle 7 parameters, one can reduce this number to 6 in a simple manner, as we now discuss. Each rotation can be written as the product of three two-dimensional rotations controlled by one parameter each,
\begin{equation}\label{Explicit f rotation}
\OLu=\roto{1}{p_{1}^{u}}\roto{2}{p_{2}^{u}}\roto{3}{p_{3}^{u}},\quad
\OLd=\roto{1}{p_{1}^{d}}\roto{2}{p_{2}^{d}}\roto{3}{p_{3}^{d}},
\end{equation}%
where each $\roto{j}{x}$ can be one of the following (with $\roto{1}{p}\neq\roto{2}{p}$, $\roto{2}{p}\neq\roto{3}{p}$)
\begin{equation}
\rot{12}{x}=\begin{pmatrix}c_x & s_x & 0\\ -s_x & c_x & 0\\ 0 & 0 & 1\end{pmatrix},\quad
\rot{13}{x}=\begin{pmatrix}c_x & 0 & s_x\\ 0 & 1 & 0\\ -s_x & 0 & c_x\end{pmatrix},\quad
\rot{23}{x}=\begin{pmatrix}1 & 0 & 0\\ 0 & c_x & s_x\\ 0 & -s_x & c_x\end{pmatrix},\quad
\end{equation}
and $c_x=\cos x$, $s_c=\sin x$.

If one chooses a parametrization with $\roto{1}{p_1^f}=\rot{12}{p_1^f}$ in both $\OLu$ and $\OLd$ (i.e. the leftmost rotation only acts in the 1-2 plane), for example
\begin{eqnarray}
\CKM &=&\OLut\phase{3}{2\theta}\OLd,  \label{VCKM} \\
\OLu &=&\rot{12}{p_{1}^{u}}\rot{23}{p_{2}^{u}}\rot{13}{p_{3}^{u}},  \label{O_u_L} \\
\OLd &=&\rot{12}{p_{1}^{d}}\rot{23}{p_{2}^{d}}\rot{13}{p_{3}^{d}},  \label{O_d_L}
\end{eqnarray}%
it is clear (note the $\phase{3}{2\theta}$ diagonal structure in \refEQS{eq:MassMatrixStruct}-\eqref{eq:PhaseProyector}) that, rather than $p_1^{u}$ and $p_1^{d}$ separately, only $p_1^{u}-p_1^{d}$ enters $\CKM$ and we can eliminate one parameter at once (we simply set $p_1^{d}=0$ without loss of generality). 
In summary, the experimental information constrains $\{\theta_{12}^q,\theta_{13}^q,\theta_{23}^q,\delta_q,\urvec{u}{1},\urvec{u}{2}\}$, and could fix the model parameters $\{p_1^{u},p_2^{u},p_3^{u},p_2^{d},p_3^{d},\theta\}$ (a full analysis along these lines was presented in \cite{Nebot:2018nqn}). The most important aspect here, in which we are interested, is the fact that, ideally, one can fix $\theta$ with this procedure, since CP violation is well established in the quark sector. 
One can address the lepton sector similarly with
\begin{eqnarray}
\PMNS &=&\OLet\phase{3}{-2\theta}\OLn,  \label{PMNS} \\
\OLn &=&\rot{12}{p_{1}^{\nu}}\rot{23}{p_{2}^{\nu}}\rot{13}{p_{3}^{\nu}},  \label{O_nu_L} \\
\OLe &=&\rot{12}{p_{1}^{e}}\rot{23}{p_{2}^{e}}\rot{13}{p_{3}^{e}}.  \label{O_e_L}
\end{eqnarray}%
Again, one can set $p_1^{\nu}=0$ without loss of generality, ending up with $\{p_1^{e},p_2^{e},p_3^{e},p_2^{\nu},p_3^{\nu},\theta\}$ as parameters in the lepton flavour sector.
The experimental information on PMNS strongly constrains $\{\theta_{12}^{\ell},\theta_{13}^{\ell},\theta_{23}^{\ell}\}$; additional information from SFCNC sensitive processes like $\nh\to \ell_i\bar\ell_j$, $i\neq j$, is needed in order to constrain or fix the parameters in \refEQ{PMNS}. The crucial point is that $\theta$ can be a priori fixed in the quark sector and thus, with one less experimental input in the leptonic sector, one could in principle predict the value of the CP violating Dirac phase $\delta_\ell$ in PMNS prior to its measurement. It is in this sense that we can ideally relate the PMNS phase to the CKM phase in this class of generalized BGL-2HDM with spontaneous CP violation.

\section{Simplified models incorporating MFV and their connections to SFCNC\label{SEC:simpMFV}}

In the full analysis of the quark sector presented in \cite{Nebot:2018nqn}, it was shown that the model was viable after surmounting a large set of constraints related to flavour transitions, Higgs signals, electroweak precision and requirements on the scalar potential. In particular, the right amount of CP violation in the CKM matrix could be accommodated, together with the presence of SFCNC. Surprisingly, this could be achieved with significant freedom in the values of $\theta$ and SFCNC. Therefore, direct generalization of the full analysis to include the lepton flavour sector does not appear to be the most promising avenue to explore the connection among CP violation in the CKM and PMNS mixing matrices in this kind of model, since this connection is blurred by this remaining freedom in $\theta$ and SFCNC.\\ 
An interesting possibility is to restrict the model by making simplifying assumptions about SFCNC, with these assumptions guided by -- and thus compatible with -- experimental data. In the following we focus on that kind of restricted scenario. 
We will first address the quark sector and then discuss the extension to the lepton sector.

%

\subsection{Quark sector\label{sSEC:QuarkSector}}

As explained in section \ref{SEC:CPSFCNC}, eliminating SFCNC either in the up or in the down quark sector, definitely simplifies SFCNC but eliminates altogether a CP violating CKM matrix. The next level of simplification would be to impose the absence of some SFCNC. Considering the general structure of SFCNC, proportional to $\urvec{f}{j}\urvec{f}{k}$, the only simplified alternative between the most general possibility and the absence of SFCNC is to assume only one vanishing component $\urvec{f}{i}=0$ for a given $i=1,2$ or $3$; in that case only one SFCNC transition $j\leftrightarrows k$ is present ($i$, $j$, $k$ all different). Notice that, a priori, one can make this kind of assumption simultaneously in both the up and the down sectors: that is, for given $i$ and $l$, one can in principle impose $\urvec{u}{i}=0$ and $\urvec{d}{l}=0$. According to the discussion in section \ref{SEC:genrel}, we have 6 parameters that need to satisfy the 4 constraints to obtain a realistic CKM matrix, to which we are now adding these 2 new requirements on SFCNC. One can indeed implement these 2 SFCNC conditions directly with a reduction from 6 to 4 parameters in our CKM matrix in \refEQ{VCKM}. Notice, in that case, that the models incorporate the MFV ansatz \cite{DAmbrosio:2002vsn,Cirigliano:2005ck}, since they have exactly as many parameters in the flavour sector as the CKM matrix requires. Furthermore, the only non-vanishing SFCNC coupling in each sector will be controlled by one of the mixing angles of CKM. As discussed in appendix \ref{SEC:MFVmodels}, one can consider, in principle, 81 different models of this type, out of which only one appears to satisfy the requirements to be viable: this model is the specific example that we consider now, step by step, in order to illustrate the central idea of the paper.
\begin{enumerate}
\item We start imposing the following form of the $\urvec{f}{}$ vectors controlling SFCNC:
\begin{equation}\label{FCtentative_Structure}
\begin{aligned}
\urvec{u}{} &=&\left( 0,-\sin p_{2}^{u},\cos p_{2}^{u}\right),\\ 
\urvec{d}{} &=&\left( -\sin p_{2}^{d},0,\cos p_{2}^{d}\right),  
\end{aligned}
\end{equation}
with parameters $p_{2}^{u}$, $p_{2}^{d}$. The assumption in \refEQ{FCtentative_Structure} is that $t\leftrightarrows c$ and $b\leftrightarrows d$ SFCNC are present while $u\leftrightarrows c$, $u\leftrightarrows t$, $d\leftrightarrows s$, $b\leftrightarrows s$ SFCNC are absent. 
\item Concerning CKM, since $\urvec{f}{i}=(\OLf{f})_{3i}$, we must have
\begin{equation}
\OLu=\rot{12}{p_1^{u}}\rot{23}{p_2^{u}},\quad \OLd=\rot{13}{p_2^d}\,.
\end{equation}
One can add a factor $\rot{12}{p_1^{d}}$ to the left of $\rot{13}{p_2^{d}}$ in $\OLd$ but, as discussed previously, this would just amount to a redefinition $p_1^{u}\mapsto p_1^{u}-p_1^{d}$. The main point is that the CKM matrix is
\begin{equation}\label{eq:CKM:MFV:00}
\CKM=\rott{23}{p_{2}^{u}}\rott{12}{p_{1}^{u}}\phase{3}{2\theta}\rot{13}{p_{2}^{d}}\,.
\end{equation}
\item Performing a fit of \refEQ{eq:CKM:MFV:00} to the measured CKM matrix (see appendix \ref{SEC:CKMPMNSfits}), one obtains
\begin{equation}\label{result of the fit quark sector}
\begin{aligned}
&2\theta=1.077^{+0.039}_{-0.031},\quad &&p_{1}^{u}=0.22694\pm 0.00052,\\
&p_{2}^{u}=(4.235\pm 0.059)\times 10^{-2},\quad &&p_{2}^{d}=(3.774\pm 0.098)\times 10^{-3}\,.
\end{aligned}
\end{equation}
In order to relate $\dCKM$ and $\dPMNS$ it is specially relevant that the quark sector fixes $\theta$.
\item In addition, \refEQ{result of the fit quark sector} fixes SFCNC with
\begin{equation}\label{SFCNC parameters from quark fit}
\urvec{u}{}=\left(0,-0.0423,0.9991\right),\quad 
\urvec{d}{}=\left( -0.0038,0,0.9999\right).
\end{equation}
A non-trivial result is that the values in \refEQ{SFCNC parameters from quark fit} are within the allowed regions arising in the analysis of \cite{Nebot:2018nqn} (for example in figures 6(b) and 6(c)). 
Even if \refEQ{SFCNC parameters from quark fit} fixes the intensity of SFCNC, the precise effects in specific processes depend on parameters such as $\tb$ and elements $\ROT{jk}$ of the mixing matrix of neutral scalars in \refEQ{eq:ScalarMass} (for example $\ROT{11}$ is the mixing among the $125$ GeV scalar and the scalar with SM Higgs couplings). With $2\theta=1.077$, $\abs{\sin 2\theta}=0.88$ and one can read in figure 9 of  \cite{Nebot:2018nqn} 
\begin{equation}\label{constraints quark sector}
\ROT{11}\in [0.82;\,0.90],\quad \tb\in [0.5,1.8].
\end{equation}
\item Then, the most relevant prediction of this model in terms of SFCNC concerns $t\to \nh c$ decays, where
\begin{equation}\label{eq:BR:thc:0}
\text{Br}(t\to c\nh) =\left( 1-\ROT{11}^{2}\right) \left(\tb+\tbinv\right)^{2}(\urvec{u}{2}\urvec{u}{3})^{2}f(x_{h},x_{W}),
\end{equation}
with
\begin{equation}\label{eq:thc:f}
f(x,y)=\frac{1}{2}\left(1-x\right)^{2}\left(1-3y^{2}+2y^{3}\right)^{-1},
\end{equation}
and $x_{i}=(m_{i}/m_{t})^{2}$, giving $f(x_{\nh},x_{W})=0.1306$ (for further details, see appendix \ref{SEC:HiggsSFCNC}). With \refEQS{SFCNC parameters from quark fit} and \eqref{eq:BR:thc:0}, one obtains
\begin{equation}\label{top to c+h BR}
1.8\times 10^{-4}\leq \text{Br}(t\to c\nh) \leq 4.3\times 10^{-4}\,.
\end{equation}
Notice that the range predicted in \refEQ{top to c+h BR} is rather reduced and indeed not far from current LHC bounds \cite{Sirunyan:2017uae,Aaboud:2018oqm}.
\item In the down sector, with $(\urvec{d}{1}\urvec{d}{3})^2\sim 10^{-2}(\urvec{u}{2}\urvec{u}{3})^{2}\sim 1.6\times 10^{-5}$, $b\leftrightarrows d$ SFCNC have a negligible effect in $B_d^0$--$\bar B_d^0$ oscillations, while $\nh\to \bar bd,b\bar d$ are beyond the LHC capabilities.
\end{enumerate}


\subsection{Lepton sector\label{sSEC:LeptonSector}}
We address now the lepton sector, applying analogous requirements on SFCNC to the ones considered for the quark sector in the previous subsection. In the lepton sector, the most stringent constraint comes from bounds on $\mu \to e+\gamma $. If we only allowed $\mu\leftrightarrows e$ SFCNC, the coupling would be controlled by $\abs{\U{ei}\U{\mu i}}^{2}$ with $i=1,2$ or $3$. Since PMNS is rather non-hierarchical, one can estimate \cite{Botella:2015hoa} that avoiding the current bound $\text{Br}(\mu\to e+\gamma)<4.2\times 10^{-13}$ \cite{TheMEG:2016wtm} requires a cancellation or fine-tuning at the $10^{-4}$--$10^{-5}$ level among scalar and pseudoscalar contributions in 2-loop Barr-Zee contributions \cite{Barr:1990vd,Chang:1993kw}. 
\begin{enumerate}
\item It is mandatory to eliminate $\mu\leftrightarrows e$ SFCNC: this can be achieved either with $\urvec{e}{1}=0$ or $\urvec{e}{2}=0$. In the neutrino sector all three choices $\urvec{\nu}{k}=0$, $k=1,2,3$ are in principle available. However, out of these 6 combined possibilities which automatically evade $\mu\to e+\gamma$ bounds, the only one which is allowed experimentally (see appendix \ref{SEC:MFVmodels} for details) is
\begin{equation}
\begin{aligned}
&\urvec{\nu}{} =\left( -\sin p_{2}^{\nu},\cos p_{2}^{\nu},0\right),\\ 
&\urvec{e}{} =\left( -\sin p_{2}^{e},0,\cos p_{2}^{e}\right). \label{FCtentative_Structure_electron}
\end{aligned}
\end{equation}%
\item With $\urvec{e}{i}=\left(\OLe\right)_{3i}$, we must have
\begin{equation}
\OLe=\rot{12}{p_{1}^{e}}\rot{13}{p_{2}^{e}}\,,
\end{equation}
and
\begin{equation}\label{eq:OLn:modelMFV}
\OLn=P_{23}\rot{12}{p_{2}^{\nu}}\,,
\end{equation}
where the permutation $P_{23}$ interchanges the third and second rows of $\rot{12}{p_{2}^{\nu}}$:
\begin{equation}
P_{23}=
\begin{pmatrix}1 & 0 & 0 \\ 0 & 0 & 1 \\ 0 & 1 & 0\end{pmatrix}\,.
\end{equation}
(Including $P_{23}$ does not depart from the general form in \refEQ{Explicit f rotation}; as explained in appendix \ref{SEC:parametrization}). As discussed previously, we do not include a left factor $\rot{12}{p_{1}^{\nu}}$ in $\OLn$, since it amounts to a redefinition of $p_1^{e}\to p_1^{e}-p_{1}^{\nu}$. 
\item The PMNS matrix is 
\begin{equation}\label{eq:PMNS:lep}
\PMNS=\rott{13}{p_{2}^{e}}\rott{12}{p_{1}^{e}}\phase{3}{-2\theta}P_{23}\rot{12}{p_{2}^{\nu}}\,.
\end{equation}
It is fully fixed by 3 mixing angles and the CP violating phase $\theta$ already obtained in the quark sector.
\item  We can fit now \refEQ{eq:PMNS:lep} to the experimental information on PMNS encoded in $\{\theta_{12}^\ell,\theta_{13}^\ell,\theta_{23}^\ell\}$ (see appendix \ref{SEC:CKMPMNSfits}). 
In this fit, $\theta$ is already set to the value obtained from the fit to the CKM matrix in \refEQ{result of the fit quark sector}. Although different PMNS analyses \cite{deSalas:2020pgw,Esteban:2020cvm} show some sensitivity to the phase $\delta_\ell$, we do not include that information in the fit since we are precisely interested in its prediction. The fit gives the following two solutions,
\begin{alignat}{4}
&\text{Solution 1}:\quad && p_{1}^{e}=0.7496,\quad &&p_{2}^{e}=1.3541,\quad && p_{2}^{\nu}=0.8974\,,\\
&\text{Solution 2}:\quad && p_{1}^{e}=2.3889,\quad &&p_{2}^{e}=1.3541,\quad && p_{2}^{\nu}=1.0542\,.
\end{alignat}
SFCNC are controlled in both cases by
\begin{equation}\label{eq:SFCNC:lep:0}
\urvec{e}{}=(-0.9765,\, 0,\, 0.2156)\,.
\end{equation}
\item Most importantly, the solutions differ in the values of the (unique) CP violating imaginary part of an invariant quartet
\begin{equation}
J_{\rm PMNS}=\im{\U{e 1}\U{\mu 2}\Uc{e 2}\Uc{\mu 1}}\,,
\end{equation}
and of the phase $\dPMNS$,
\begin{equation}
\begin{tabular}{|c|c|c|c|c|}
\hline
Case & $J_{\rm PMNS}$ & $\dPMNS$ & $\Delta\chi^2_{\rm NO}(\dPMNS)$ & $\Delta\chi^2_{\rm IO}(\dPMNS)$\\ \hline
\text{Solution 1} & $-0.0316$ & $1.629\pi\,(293^\circ)$ & $5$ & $0$\\ \hline
\text{Solution 2} & $0.0282$ & $0.679\pi\,(126^\circ)$ & $13$ & $>20$\\ \hline
\end{tabular}
\end{equation}
$\Delta\chi^2_{\rm NO}(\dPMNS)$ and $\Delta\chi^2_{\rm IO}(\dPMNS)$ show the values that correspond to $\dPMNS$ attending to the $\Delta\chi^2$ profiles for $\delta_\ell$ obtained for normal and inverted neutrino mass orderings in \cite{deSalas:2020pgw}.\\ 
We stress that using the information on CP violation in the quark sector, we have been able to predict the phase in PMNS using the connection that SCPV provides in this model; in particular, Solution 1 has $\dPMNS=1.629\pi$, which is in good agreement with the most likely values in PMNS analyses.
\item With the values of $e\leftrightarrows\tau$ SFCNC in \refEQ{eq:SFCNC:lep:0}, we have predictions (see appendix \ref{SEC:HiggsSFCNC}) such as
\begin{equation}\label{eq:BRhetau}
\text{Br}(\nh\to e\bar\tau+\bar e\tau)=
\left( 1-\ROT{11}^{2}\right)\left(\tb+\tbinv\right)^{2}(\urvec{e}{1}\urvec{e}{3})^{2}\,\text{Br}(\nhSM\to \tau\bar\tau)
\,\frac{\Gamma(\nhSM)}{\Gamma(\nh)} \,.
\end{equation}
Using \refEQ{constraints quark sector}, we have the sharp range
\begin{equation}
2.0\times 10^{-3}\leq\text{Br}(\nh\to e\bar\tau+\bar e\tau)\frac{\Gamma(\nh)}{\Gamma(\nhSM)}\leq 5.0\times 10^{-3},
\end{equation} 
which should be seen or disproved in the near future since the current bound is\footnote{Recent work \cite{Sirunyan:2021ovv} might lower this bound closer to $2\times 10^{-3}$.} $\text{Br}(\nh\to e\bar\tau+\bar e\tau)_{\rm Exp}\leq 4.7\times 10^{-3}$ \cite{Sirunyan:2017xzt,Aad:2019ugc,Davidek:2020gbw} (although there is some freedom in ${\Gamma(\nh)}/{\Gamma(\nhSM)}$, it does not modify substantially this conclusion).
\end{enumerate}

\section*{Conclusions\label{SEC:Concl}}
We have discussed the possibility of having a framework where there is a connection between the CP violations in the quark and the lepton sectors, parametrised by the two phases $\dCKM$ and $\dPMNS$. In general, it is not possible to establish this connection, since the Yukawa couplings $Y_{f}$ -- with $f=u,d,\ell,\nu$ -- generating the quark and lepton masses are complex matrices, with no relation between $Y_{u,d}$ and $Y_{\ell,\nu}$. In this paper we have investigated this connection in a framework where CP violation in the quark and lepton sectors have a common origin, being both generated by a complex vacuum phase. We have pointed out that in order to construct experimentally viable models of this class, some conditions have to be satisfied, namely the vacuum phase has to be able to generate complex CKM and PMNS matrices. This is not an easy task since it is assumed that CP is spontaneously broken, so the Yukawa couplings are real.
We have shown that in order to generate a complex CKM matrix, one has to have SFCNC both in the up and down quark sectors. An entirely analogous requirement applies to the lepton sector, where the generation of a complex PMNS matrix also requires the presence of scalar leptonic FCNC.
Since there are stringent bounds on these SFCNC, we consider generalised BGL 2HDMs where there is natural suppression of some of the most dangerous SFCNC. We have shown that within some of the generalised BGL models, there is indeed a connection between $\dCKM$ and $\dPMNS$.
The interplay among CPV and the existence of SFCNC makes that these relations are quite involved implying connections or predictions for processes mediated by SFCNC in all the sectors: up, down quarks and charged leptons\footnote{SFCNC involving neutrinos, being proportional to the neutrino masses are not experimentally accessible.}. To clarify all these relations we have worked a subclass of our most general model of SCPV, but guided by two important hints: (i) an experimental fact: the absence of any convincing evidence of the presence of SFCNC at the actual level of precision; (ii) a theoretical discovery: the necessity of having at least one type of SFCNC in each sector: quarks up and down, charged leptons and even in the neutrino sector.
Therefore we have worked with models that have the minimal amount of SFCNC needed to keep SCPV. These simplified models verify the MFV ansatz. Because they are controlled by the four unit vectors $\urvec{u}{}$, $\urvec{d}{}$, $\urvec{\nu}{}$, $\urvec{e}{}$ having a zero in some entry, there are $3^4=81$ possible models of this type. In the model discussed in section \ref{SEC:simpMFV}, the connection between $\dCKM$ and $\dPMNS$ gives a prediction for $\dPMNS$ in agreement with recent PMNS analyses. In the appendices we explain how the remaining models confront experimental data requirements.

\section*{Acknowledgments}
The authors acknowledge support from \emph{Funda\c{c}\~ao para a Ci\^encia e a Tecnologia} (FCT, Portugal) through the projects CFTP-FCT Unit 777 (UIDB/00777/2020 and UIDP/00777/2020), PTDC/FIS-PAR/29436/2017 and CERN/FIS-PAR/0008/2019, which are partially funded through POCTI (FEDER), COMPETE, QREN and EU, from Spanish grants FPA2017-85140-C3-3-P and FPA2017-84543-P (AEI/FEDER, UE) and from \emph{Generalitat Valenciana}, Spain, under grant PROMETEO 2019-113. 
FCG is supported by \emph{Ministerio de Ciencia, Innovación y Universidades}, Spain, under grant BES-2017-080070 and partially supported by a Short-Term Scientific Mission Grant from the COST Action CA15108.
MN is supported by the \emph{GenT Plan} from \emph{Generalitat Valenciana} under project CIDEGENT/2019/024.

\appendix

\clearpage
\section{Higgs SFCNC\label{SEC:HiggsSFCNC}}
The Yukawa couplings of the Higgs-like scalar $\nh$, following section \ref{SEC:Model}, read
\begin{equation}\label{eq:HiggsYuk}
\mathscr L_{\nh f\bar f}=-\frac{\nh}{\vev{}}\bar f_{j}\left\{\left[\ROT{11}m_{f_j}\delta_{jk}+\ROT{21}H^f_{jk}\pm i\ROT{31}A^f_{jk}\right]+\left[\ROT{21}A^f_{jk}\pm i\ROT{31}H^f_{jk}\right]\gamma_5\right\}f_k
\end{equation}
where
\begin{eqnarray}
H^f_{jk}&=&\tb\,m_{f_j}\delta_{jk}-(\tb+\tbinv)\urvec{f}{j}\urvec{f}{k}\frac{m_{f_j}+m_{f_k}}{2},\\
A^f_{jk}&=&(\tb+\tbinv)\urvec{f}{j}\urvec{f}{k}\frac{m_{f_j}-m_{f_k}}{2}\,.
\end{eqnarray}
The sign $\pm$ appearing with $\ROT{3s}$ is $+$ for $f=d,\ell$ and $-$ for $f=u,\nu$. One can immediately read the SFCNC terms relevant for $\nh\to e\tau$ (neglecting $m_e$ terms)
\begin{equation}
\mathscr L_{\nh e\tau}=\frac{m_\tau(\tb+\tbinv)\urvec{e}{1}\urvec{e}{3}(\ROT{21}-i\ROT{31})}{2\vev{}}\,\nh\bar e(1+\gamma_5)\tau+\text{h.c.}.
\end{equation}
In the SM, the coupling of $\nhSM$ to $\bar\tau\tau$ is simply $-\frac{m_\tau}{\vev{}}\nhSM\bar\tau\tau$ and, with $|\ROT{21}-i\ROT{31}|^2=1-\ROT{11}^2$, one can obtain straightforwardly \refEQ{eq:BRhetau}.
Similarly, for $t\to\nh c$ decays (neglecting $m_c$ terms) the interaction term is
\begin{equation}\label{eq:SFCNC:thc}
\mathscr L_{\nh \bar ct}=\frac{m_t(\tb+\tbinv)\urvec{u}{2}\urvec{u}{3}(\ROT{21}+i\ROT{31})}{2\vev{}}\,\nh\bar c(1+\gamma_5)t\,,
\end{equation}
and one obtains $\text{Br}(t\to c\nh)$ in \refEQ{eq:BR:thc:0} considering that the top quark width is dominated by $t\to Wb$, i.e. $\Gamma(t)=\Gamma(t\to Wb)$ with $\abs{\V{tb}}\simeq 1$; $f(x_{\nh},x_{W})$ in \refEQ{eq:thc:f} collects the differences among both decays due to (i) scalar $\nh$ vs. vector $W$ in the final state and (ii) $\nh c$ vs. $Wb$ phase space.

\section{CKM and PMNS fits\label{SEC:CKMPMNSfits}}
The PDG parametrization \cite{Chau:1984fp,Zyla:2020zbs} of the CKM and PMNS matrices is
\begin{multline}\label{eq:PDG:unit33}
\rot{23}{\theta_{23}}\phase{3}{\delta}\rot{13}{\theta_{13}}\phase{3}{-\delta}\rot{12}{\theta_{12}}=\\
\begin{pmatrix}
c_{12}c_{13} & s_{12}c_{13} & s_{13}e^{-i\delta}\\
-s_{12}c_{23}-c_{12}s_{13}s_{23}e^{i\delta} & c_{12}c_{23}-s_{12}s_{13}s_{23}e^{i\delta} & c_{13}s_{23}\\
s_{12}s_{23}-c_{12}s_{13}c_{23}e^{i\delta} & -c_{12}s_{23}-s_{12}s_{13}c_{23}e^{i\delta} & c_{13}c_{23}
\end{pmatrix},
\end{multline}
where rephasings allow to reduce the parameter ranges to $\theta_{ij}\in[0;\pi/2]$ and $\delta\in[0;2\pi[$.\\  %
For CKM we use the values \cite{Zyla:2020zbs}
\begin{equation}\label{eq:CKMinputs}
\begin{aligned}
&\sin\theta_{12}^q=0.2265\pm 0.0005\,,\quad &&\sin\theta_{13}^q=(3.61\pm 0.10)\times 10^{-3}\,,\\
&\sin\theta_{23}^q=(4.05\pm 0.07)\times 10^{-2}\,,\quad && \delta_q=(66.9\pm 2.0)^\circ\,.
\end{aligned}
\end{equation}
For the PMNS matrix, we use
\begin{equation}\label{eq:PMNSinputs}
\sin^2\theta_{12}^\ell=0.32^{+0.020}_{-0.016}\,,\quad \sin^2\theta_{13}^\ell=(2.16^{+0.063}_{-0.066})\times 10^{-2}\,,\quad \sin^2\theta_{23}^\ell=0.547^{+0.020}_{-0.030}\,.
\end{equation}
Although results corresponding to normal ordering and inverted ordering of neutrino masses differ slightly, and results quoted by several groups differ too \cite{Zyla:2020zbs,deSalas:2020pgw,Esteban:2020cvm}, these differences are unsubstantial for the scope of this work.\\ 
To illustrate how one fits the CKM and PMNS matrices, in the models under consideration, to the experimental information condensed in \refEQS{eq:CKMinputs} and \eqref{eq:PMNSinputs}, let us consider the CKM case. In terms of the parameters of the particular model under consideration, a CKM matrix $\CKM$ is computed. Then, one computes (for simplicity, we use here $1,2,3$ indices rather than $u,c,t,d,s,b$)
\begin{equation}
s_{13}^2=\abs{\V{13}}^2,\qquad s_{12}^2=\frac{\abs{\V{12}}^2}{1-\abs{\V{13}}^2},\qquad s_{23}^2=\frac{\abs{\V{23}}^2}{1-\abs{\V{13}}^2},
\end{equation}
and
\begin{equation}
\delta=\arg\left(\frac{\V{12}\Vc{13}\V{23}\Vc{22}}{\abs{\V{12}\V{13}\V{23}}(1-\abs{\V{13}}^2)}+\abs{\V{12}\V{13}\V{23}}\right)
\end{equation}
These are the values, obtained in a rephasing invariant manner, to be compared with \refEQ{eq:CKMinputs} (in the actual fit, this comparison gives the usual likelihood function of the model parameters, which is maximized). For the PMNS matrix, we follow the same procedure with the values in \refEQ{eq:PMNSinputs}, and include no constraint on the phase $\delta_\ell$, as discussed in section \ref{SEC:simpMFV}. 
 
\section{MFV models\label{SEC:MFVmodels}}
In section \ref{SEC:simpMFV} we have introduced a class of models in which SFCNC are only present in a single transition in the up and down quark sectors: that is the minimal assumption which can produce a complex CKM matrix. This assumption is then extended to the lepton sector. One can consider, in principle, $3^4=81$ different models in this class (3 choices for the vanishing component of $\urvec{f}{}$ for $f=u,d,\nu,e$). In the lepton sector, however, following the considerations on $\mu\leftrightarrows e$, we directly discard models with $\urvec{e}{3}=0$. In the following we analyse how there is only one model which satisfies some basic requirements, the model considered in section \ref{SEC:simpMFV}.\\ 
We use the following notation to identify the different models: model $(ud)=(jk)$, $j,k=1,2,3$, is the model in which $\urvec{u}{j}=0$ and $\urvec{d}{k}=0$: $j$ identifies which generation does not have SFCNC in the up quark sector and similarly for $k$ in the down quark sector. For the quark sector, the model in section \ref{SEC:simpMFV} with $\urvec{u}{}$ and $\urvec{d}{}$ in \refEQ{FCtentative_Structure} is model $(ud)=(12)$. The extension of the notation to the lepton sector is straightforward: with $\urvec{\nu}{}$, $\urvec{e}{}$ in \refEQ{FCtentative_Structure_electron}, the complete label of the model is $(ud,\nu\ell)=(12,32)$.\\
In principle, one would need to perform an analysis of each model similar to the one in \cite{Nebot:2018nqn}, extended to the lepton sector, a task which is beyond the scope of this work. It is nevertheless possible to consider simpler analyses and a few requirements to understand how all but one model have to be discarded.\\
The first requirement in the quark sector is that a CKM matrix in agreement with data can be obtained. All 9 models $(ud)=(jk)$, $j,k=1,2,3$, can give a good CKM matrix. The next requirement concerns SFCNC: if one sorts the absolute value of the components of $\urvec{f}{}$, $\abs{\urvec{f}{Max}}\geq\abs{\urvec{f}{Mid}}\geq\abs{\urvec{f}{Min}}$, the presence of SFCNC is necessary in order to obtain a complex CKM matrix, which means $\abs{\urvec{f}{Mid}}>0$. Notice that in the class of models under consideration, $\urvec{f}{Min}=0$. In the analysis of \cite{Nebot:2018nqn}, rather restrictive ranges for $\abs{\urvec{u}{Mid}}$ and $\abs{\urvec{d}{Mid}}$ were obtained. We impose that constraint as a proxy for more involved analyses, explicitely we require
\begin{equation}\label{eq:rMid:ranges}
\abs{\urvec{u}{Mid}}\in [0.04; 0.25],\quad \abs{\urvec{d}{Mid}}\in [0.003;0.10].
\end{equation}
Models $(ud)=(13)$, $(21)$, $(22)$, $(23)$, $(33)$, violate grossly that requirement when a good CKM matrix is obtained. Although model $(ud)=(11)$ can give values close to the ranges in \refEQ{eq:rMid:ranges}, it cannot produce a good CKM matrix while  satisfying \refEQ{eq:rMid:ranges}. Models $(ud)=(31)$ and $(32)$ can produce a good CKM matrix and also fulfill \refEQ{eq:rMid:ranges} with $\abs{\urvec{u}{1}\urvec{u}{2}}\simeq 0.22$. The corresponding $\nh$-SCFNC interaction, analogous to \refEQ{eq:SFCNC:thc}, is of the form $\mathscr L_{\nh uc}=C_{ud}^{\nh}\overline{c_L}u_R\nh$. It gives rise, at tree level, to an effective operator $\frac{(C_{ud}^{\nh})^2}{\mnh^2}(\overline{c_L}u_R)$ contributing to $D^0$--$\bar D^0$ mixing. Omitting the possibility of significant cancellations with similar contributions mediated by $\nH$ and $\nA$, $D^0$--$\bar D^0$ mixing sets strong bounds on $C_{ud}^{\nh}$ (see for example \cite{Blankenburg:2012ex}), which cannot be satisfied within models $(ud)=(31)$ and $(32)$.\\ %
The only model left in the quark sector is $(ud)=(12)$. The problem of exploring the viable models is then reduced to analyse the requirements on the lepton sector of the 6 models $(ud,\nu\ell)=(12,jk)$, $j=1,2,3$, $k=1,2$. Models $(\nu\ell)=(11)$, $(21)$, $(31)$, can produce a realistic PMNS matrix, but then $\abs{\urvec{e}{2}\urvec{e}{3}}$ is close to its maximal value, $1/2$, yielding too large $\text{Br}(\nh\to\mu\tau)$ predictions. On the other hand, models $(\nu\ell)=(12)$, $(22)$, cannot produce a realistic PMNS matrix; model $(\nu\ell)=(32)$, on the contrary, can produce a good PMNS matrix. We are left with $(ud,\nu\ell)=(12,32)$, the model considered in section \ref{SEC:simpMFV}, as the only viable candidate.
 
\section{Parametrizations\label{SEC:parametrization}}
In section \ref{sSEC:LeptonSector}, \refEQ{eq:OLn:modelMFV}, a permutation $P_{23}$ is introduced in the parametrization of $\OLn$ which deviates, apparently, from the general parametrization introduced in \refEQ{Explicit f rotation}. While $\rot{13}{p}$ and $\rot{23}{p}$ naturally have one vanishing component in the third row (the one controlling SFCNC) in position 2 and 1 respectively, parametrizing the MFV models introduced in section \ref{SEC:simpMFV} in which the vanishing component is in position 3 requires an additional consideration. This permutation $P_{23}$ is introduced in order to properly implement that case. One can indeed rewrite
\begin{multline}
P_{23}=\begin{pmatrix}1&0&0\\ 0&-1&0\\ 0&0&1\end{pmatrix}\begin{pmatrix}1&0&0\\ 0&0&-1\\ 0&1&0\end{pmatrix}=\begin{pmatrix}1&0&0\\ 0&0&-1\\ 0&1&0\end{pmatrix}\begin{pmatrix}1&0&0\\ 0&1&0\\ 0&0&-1\end{pmatrix}=\\
\text{diag}(1,-1,1)\,\rot{23}{\pi/2}=\rot{23}{\pi/2}\,\text{diag}(1,1-,1),
\end{multline}
that is, $P_{23}$ can be simply viewed as a product of a fixed rephasing and a 2-3 rotation with fixed angle $\pi/2$ and thus there is nothing essentially different with respect to the other cases.\\ 
It is to be mentioned that while the PDG parametrization of CKM and PMNS in appendix \ref{SEC:CKMPMNSfits} is such that one can reduce the ranges of $\theta_{ij}$ to $[0;\pi/2]$ (while $\delta\in[0;2\pi[$) through rephasings, in our case the ranges of the parameters $p_1^f$, $p_2^f$ in the different orthogonal matrices $\OLf{f}$ entering CKM and PMNS, require some care since they cannot be completely reduced to those ranges (the form of $\CKM$ and $\PMNS$ in \refEQS{CKM and PMNS}, \eqref{VCKM}-\eqref{PMNS}, is different from \refEQ{eq:PDG:unit33}).


\providecommand{\href}[2]{#2}\begingroup\raggedright
\endgroup
\end{document}